\documentclass[11pt,twoside]{article}

\usepackage{asp2010}

\resetcounters

\begin{document}
\title{Teaching Optics and Systems Engineering With Adaptive Optics Workbenches}   
\author{D.~M.~Harrington$^1$, M.~Ammons$^2$, Lisa~Hunter$^3$, Claire~Max$^4$, Mark~Hoffmann$^5$, Mark~Pitts$^1$, J.~D.~Armstrong$^6$} 

\affil{$^1$Institute for Astronomy, University of Hawai`i, Honolulu, HI 96822} 
\affil{$^2$Laboratory for Adaptive Optics, University of California, Santa Cruz, CA, 95064}
\affil{$^3$Institute for Science and Engineer Educators, University of California, Santa Cruz, CA, 95064}
\affil{$^4$Center for Adaptive Optics, University of California, Santa Cruz, CA,  95064}
\affil{$^5$University of Hawai`i, Maui College, Kahului, HI, 96732}
\affil{$^6$Institute for Astronomy Maui, University of Hawai`i, Pukalani, HI 96768}

\begin{abstract} Adaptive optics workbenches are fully functional optical systems that can be used to illustrate and teach a variety of concepts and cognitive processes. Four systems have been funded, designed and constructed by various institutions and people as part of education programs associated with the Center for Adaptive Optics, the Professional Development Program and the Institute for Science and Engineer Educators. Activities can range from first-year undergraduate explorations to professional level training. These workbenches have been used in many venues including the Center for Adaptive Optics AO Summer School, the Maui Community College-hosted Akamai Maui Short Course, classrooms, training of new staff in laboratories and other venues. The activity content has focused on various elements of systems thinking, characterization, feedback and system control, basic optics and optical alignment as well as advanced topics such as phase conjugation, wave-front sensing and correction concepts, and system design. The workbenches have slightly different designs and performance capabilities. We describe here outlines for several activities utilizing these different designs and some examples of common student learner outcomes and experiences.
\end{abstract}

\section{Introduction}

	 In order to effectively illustrate and teach optics knowledge as well as train optical engineering process skills, hands-on experience and investigations using sophisticated optical hardware is critical. With this in mind, a large group of Center for Adaptive Optics (CfAO) and Institute for Scientist and Engineer Educators (ISEE) affiliated educators have designed, built, and developed adaptive optics (AO) workbenches for use in the classroom and in the laboratory. When teaching the concepts and the technology to audiences such as the public, undergraduate and graduate students or professionals, having a functioning AO system allows for a wide range of teaching methods to be utilized to engage an audience at many skill levels. We aim here to describe these activities in some detail with as general a language as possible so that elements of the activities may be adapted to other settings.

\subsection{Adaptive Optics and the Workbenches}

	 Adaptive optics in a broad sense are optical elements that can change their shape in a controlled way. There are many different technologies involved but a common adaptive device in astronomy, vision science and communications is a mirror that deforms its shape in order to correct optical systems. These deformable mirrors (DMs) are almost always flat surfaces that bend or curve when elements on the back of the mirror (called actuators) apply forces to the mirror. There are many variations on AO system design and performance depending on whether they are correcting starlight on large telescopes, imaging the retina through the blurring effects of the eye, or being used for controlling the spatial properties of a beam of light.

\begin{figure} [!h, !t, !b]
\begin{center}
\includegraphics[width=0.85\linewidth]{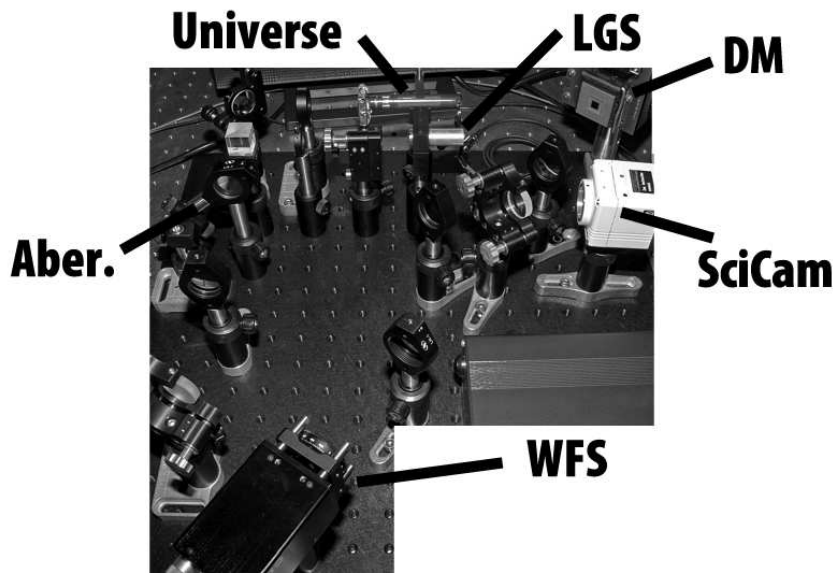}
\label{ifa}
\caption{The AO workbench constructed at the Maui IfA. A laser guide star (LGS) and star scene (Universe) are combined and aberrated (Aber.). The deformable mirror (DM) and wavefront sensor (WFS) collectively correct the image that is seen by the science camera (SciCam). See the text for details.}
\end{center}
\end{figure}

	 Though the technology involved in real AO systems can be wide ranging, the underlying concepts are essentially the same. The concepts range across disciplines such as geometric and physical optics or systems engineering but can be taught effectively by using various activity types. An example of an AO workbench constructed at the Maui Institute for Astronomy (IfA) is shown in Figure 1. The workbench is essentially optics that first create a blurred beam of light that looks like what an astronomical telescope would see and then corrects this blur using adaptive optics. The AO workbench is effectively three sub-systems: One that creates and blurs an image, one that measures the blurring and one that corrects the blurring. These sub-systems can be called the simulator (or aberrator), the wavefront sensor (WFS), and wavefront corrector (or some may simply say the DM). A laser guide star or LGS is combined with some simulated astronomical star scene and a blurring optic called an aberrator is inserted to mimic the effects of the atmosphere on light from a distant astronomical source. The blur caused by this aberrator is sensed by a lens-array and detector. The sensed blurring is then removed by a deformable mirror. The wavefront sensor and wavefront corrector must communicate via a control loop and the performance of this control system also influences the system performance.

	There are many opportunities for teaching the optical concepts related to the details of the three subsystems (aberration, sensing, correcting) as well as basic optics and engineering. The actual design and capability of the workbenches will influence the activities outlined here and the workbenches themselves can be adapted to suit educational needs.

\subsection{History of AO Workbench Activities}

	There are presently four AO workbenches that function in the various educational activities affiliated with CfAO, ISEE, IfA, Maui Community College (MCC), and Hawai`i Community College (HCC). Many instructors have used them in activities related to their Professional Development Program teaching duties (see Hunter et al., this volume). The first workbench was built by Mark Ammons in 2006 while at the Lab for Adaptive Optics (LAO) at UC Santa Cruz with help from Claire Max. A second AO workbench was developed at HCC by James Ah Heong mentored by Mark Ammons in conjunction with the Iris AO company Small Business Integrated Research program (SBIR), and with assistance from the National Science Foundation (NSF) Research Experience for Undergraduate (REU) program. As educational collaborations increased scope on Maui with MCC and their Electronic Computing and Engineering Technology (ECET) program another AO workbench was funded and developed for MCC courses and CfAO affiliated summer programs. Joe Curamen, mentored by Mark Ammons and Mark Hoffman, built the system. Finally, in the summer of 2008, the IfA Maui funded the development of a fourth system with Iris AO and the NSF REU program. Darcy Bibbs was mentored both at IfA Maui by Dave Harrington and was hosted at LAO and mentored by Iris AO, Mark Ammons, Lisa Hunter and Scott Seagroves.

	These four AO workbenches have been used at many venues over the last several years. The CfAO workbench participates in the annual CfAO Adaptive Optics Summer School (AOSS) described by Ammons et al.\ (this volume).  Graduate classes and graduate student training on adaptive optics at CfAO utilize the workbenches. The HCC workbench has also participated in the AOSS as well as HCC classes and public demonstrations.  The MCC workbench is used as part of the ECET programs undergraduate course work. In the Akamai Workforce Initiative (AWI) annual summer internship programs short-course (the Akamai Maui Short Course or AMSC) the workbenches are used in an all-day activity described in Montgomery et al.\ (this volume). The IfA Maui workbench is used in the CfAO AOSS and public demonstrations. MCC is transitioning to a four-year college: University of Hawai`i, Maui College (UHMC). The new IfA Maui workbench will be used heavily in the new undergraduate classes being taught at IfA Maui as part of the new four-year Engineering Technology degree at MCC.

\section{AO Workbench Function and Capability}

	There are different designs for AO workbenches depending on the budget for the activity, the size and capability of the DM, the type of aberration corrected and the optical or engineering concepts highlighted in the activity. The three main subsystems can be treated fairly independently and there is one control loop between the DM and the WFS. In order to understand the design choices and trade-offs in AO workbench performance, a clear listing of the critical functions and optics must be made. With a broad audience of educators in mind, we will outline the core concepts and functions in full.

\subsection{Optical Layout}

	The most simple AO workbench design would start with a laser as a collimated light source (with the light all traveling in a straight line, rays parallel).  Some source of aberration would then be inserted after the laser. This aberrator could be as simple as a clear piece of plastic, a low-powered lens or some glass with a thin coat of hairspray. The aberrator could also be as precise as a phase screen from an optical manufacturer designed to produce a specific power spectrum of wave-front errors in a motorized rotation stage. 

	The next set of components is a pair of lenses picked to expand or compress the laser beam to be the same size as the DM. Almost always the DM is bigger than the laser so we can call this pair of lenses the beam expander. The lenses of the beam expander must be `confocal' meaning that they both are mounted so that collimated light coming in also leaves the lenses collimated. They both share a common focal point in between them.
	
	As a confocal pair of lenses, this beam expander creates an image of an object placed inside the focus of the first lens somewhere after the beam expander. For instance, if the aberrator is paced one focal length away from the first lens, the aberrator is imaged by the beam expander at one focal length away from the second lens. The beam expander is then said to relay the image of the aberrator downstream and the lens pairs are also called relays. 
	
	The next element in an AO workbench is typically the DM, which is mounted in the beam exactly where the aberrator image is relayed. This is called being optically conjugate. Restated, the beam expander re-images the aberrator onto the DM (magnified), and the aberrator and DM are optically conjugate. 

	The next element is another beam expander which re-images (relays) the DM onto the final optical element -- the lenslet array (LLA). This LLA is the key concept in the wavefront sensor. The lenslet array is a single optical element that is a grid of small lenses. These LLAs are available from many optics manufacturers. 

	Most workbench designs include a method for recording the uncorrected and corrected beams. The simplest method to achieve this is to mount a beam splitter (BS) in the middle of the second beam expander. This BS sends half the light in one direction and half the light in another direction. One path of the light is sent to a detector placed at the focus of the first beam-expander lens. The other light path continues through the second beam expander lens and continues on to the LLA. This beam splitter and detector system is typically called the science camera because this camera records the corrected beam and is the desired product of AO systems. The science camera gives a means for visually inspecting the correction and demonstrating the system function as well as a method to record images and characterize how well the image is corrected. A conceptual layout of this AO workbench is shown in Figure 2. 

	In summary, first a beam is created and aberrated to simulate a distorted input. The aberrator is relayed on to the DM which is also relayed on to the LLA. The aberrator, DM and LLA are all optically conjugate. Two confocal lens pairs (relays) also serve to magnify and re-image these three optical elements. There is a BS which gives access to the corrected beam at the science camera.

\subsection{Software \& DM Control}

\begin{wrapfigure}{r}{0.6\textwidth}
\includegraphics[height=0.65\textwidth, angle=0]{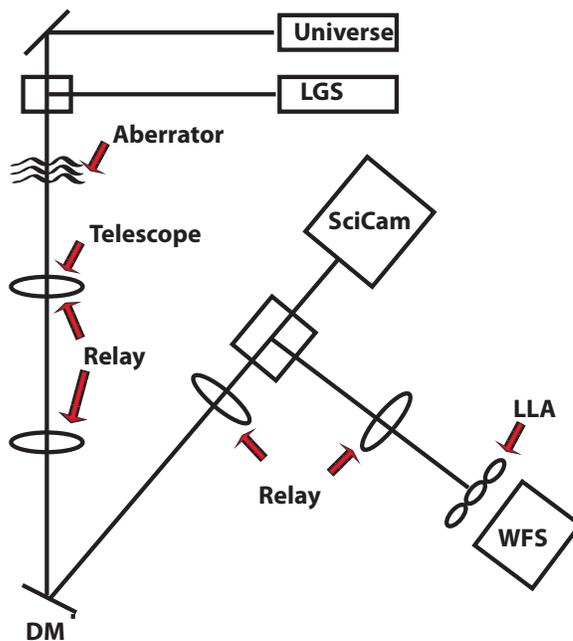}
\caption{\label{layout} Conceptual layout for an AO demonstrator}
\end{wrapfigure}

	Once the optical layout is complete, one must choose how to implement the wave-front sensing and the software control of the DM. There are many design choices but the two main designs implemented for the AO workbenches are discussed here. All four AO workbenches use a Shack-Hartmann (S-H) type WFS which is effectively a sensor system that digitally records the LLA spots, finds the center of the spots and somehow calculates a wave-front shape from those spots. There are effectively two different designs for the AO workbenches.

	There are two different DM and WFS types which come with two different software control interfaces. The CfAO and MCC workbenches use A-Optics DMs of the drum head type. They are about 8mm in diameter and are electrostatically controlled. These mirrors are good for correcting wave-fronts that look out of focus. However the DM cannot take more complicated shapes easily. The IrisAO DMs are segmented, about 4mm in diameter and can take complicated shapes rapidly. 
	
	The CfAO and MCC workbench software was written by Mark Ammons to do the WFS image analysis and to control the A-Optics mirrors. There are programs to apply voltages and to twitch the various actuators on the DM. The HCC and IfA workbenches use an IrisAO software package that works with a specific LLA and detector. The package comes with programs to apply various shapes to the DM and to twitch any segment of the DM.

\section{Activities}

	The activity types are heavily dependent on the content and cognitive processes to be taught. The main optical content areas easily covered with a system like this are optical alignment, collimation, beam expansion and compression, re-imaging, optical relays, phase conjugation, aberration, wavefronts, wavefront sensing and beam correction. The main systems engineering content areas that can be covered are system control, feedback loops, system performance, system characterization and for more in depth activities, systems optimization. Technical processes that can be reinforced include optical alignment techniques, systems thinking, block-diagramming and systems integration.

	The types of educational activities that can be performed with a system like this range from short undergraduate-level investigations to very thorough, highly technical professional-level engineering challenges.

\subsection{Understanding and Outlining Systems}

	In a short activity taught at the AMSC and as a warm up to a more thorough investigation at the AOSS, students who are already familiar with some AO terminology can do a simple systems thinking / block diagramming exercise. Depending on the background level of the students, an exercise can be given to identify components by name, draw a block diagram for the system, draw a complete optical ray trace of the system and create a `conceptual' ray trace of the system. Students are simply shown the system functioning with a minimal amount of explanation and are asked to trace light throughout the system and to accomplish their task by drawing diagrams of the system as a team. 
	
	Diagramming as a skill can be facilitated by encouraging group discussion of observations of the light path and the detector displays. By tracing the light with a notecard, students identify where the beam is collimated, where there are focal points in the beam, where the beam diameters change and which detectors display what kind of image. Since the LLA produces an array of spots on one sensor while the science camera outputs an image of an astronomical object, students can identify the main light pathways and the main optics for each system. The computer has a program which tracks the LLA spots and the spots move substantially when the aberrator is inserted. Since the main required output is a diagram, a group discussion over good diagram properties, physical layouts versus conceptual block-diagrams and clear communication can be scaffolded.
	
	Students can be assessed on the correctness of their parts identification, the accuracy of their diagrams and the completeness of their ray-traces.  This simple exercise reinforces the concepts of beam expansion and compression, collimation and allows students to outline the sub-systems in a workbench. Since this is a short, guided activity, facilitating group discussion and diagramming are important strategies to engage learners in relating their observations and drawings to other students. This activity is a first step in understanding the functioning of the workbench and can also serve as an introductory activity as part of a more in-depth longer investigation.

\subsection{Systems Thinking and Optical Concepts}

	In the AMSC, the above type of systems thinking exercise was used in a three-hour investigation. The goals of the AMSC exercise were to give students an opportunity to apply some of their previous knowledge to understand the functions of the AO workbench subsystems. The activity is more thoroughly described in Montgomery et al.\ (this volume).

	The activity starts with an introductory talk to motivate AO by showing results and (potentially) introducing terminology the instructor wants to investigate. Depending on the background level of the students this can be minimal to substantial. The three stations at the AMSC corresponded to 1) wavefront sensing 2) wavefront correction and 3) systems characterization.  
	
	In the wavefront sensing station, students investigate the relationship between the incidence angle of rays hitting a lens and the location of the corresponding image. The students measure tilt angles and focal point deviations to establish a measured relationship.  The students then must `invert' this relationship to show how a measured focus location can be used to infer the average wavefront tilt. In one implementation of this station, students used their relation to predict the angle of a mirror inside a box by setting up a lens referenced to their light source and computing the angle from a measured focus. Students must measure and document a relationship for a specific lens the apply their knowledge of this lens to infer the orientation of an external mirror. 
	
	The wavefront correcting station consisted of large plastic fresnel lenses on a table with a light source and mirrors.  A lamp was collimated by a lens and reflected off a curved mirror to aberrate the beam. This aberrating mirror was mostly hidden by a cardboard box so that the shape and effect was an unknown. A flat mirror could be inserted in front of the aberrator box to bypass the blurring effect of the aberrator. This aberrated beam was then focused and re-collimated by two more lenses. Another mirror was mounted to reflect the beam towards a final lens which formed an image on a screen. The aberrated image of the lamp was noticeably blurred while the unaberrated beam was a clear image. Students are told that the lamp and first lens simulate a star and that the `aberrator box' simulated the blurring of the atmosphere. The engineering task was to use the provided materials (foil, plastic, flat mirrors, bendable mirrors, and more lenses) to correct the beam.  Students had access to both an intermediate image as well as a point optically conjugate to the aberrator where a bendable mirror could perform the same correction as a DM. 
	
	This engineering challenge fostered group discussions of optical concepts like collimation, image formation, image sharpness and phase conjugation. The students must apply their optics knowledge to align and focus the lenses and then to determine the effects of various materials on a beam. This explores the concepts of aberration and wavefront distortion. 
	
	The systems characterization station was done using the AO workbench. Students did the short system outline activity described above. They were shown the functioning system and asked to trace the light and identify what components performed which functions. If students had taken optics classes, they were also facilitated towards drawing an accurate ray-trace of the system. A few students were also able to take their ray-trace and create a `conceptual' ray trace to stimulate discussion of beam compressors and optical conjugation. Groups were facilitated toward discussions of block diagrams and identifying sub-system functions and outputs.

\subsection{Systems Design and Optical Conjugation}

	An activity at the CfAO AO Summer School expanded on the optical content and technical skills.  The AOSS is a summer school for graduate students, college faculty and industry professionals on adaptive optics described in Ammons et al.\ (this volume). The lab section of the summer school is typically three separate labs on the second or third day of a five-day course. The participants have a wide background of optics experience ranging from exposure to optics in undergraduate physics classes to multiple years doing optics in a laboratory setting. Some are even AO specialists though many were more experienced at AO operation and data analysis instead of system design and construction.
	
	The design of the AOSS activity was much more focused on hands-on experience and on the optical concepts of conjugation and systems performance. The activity was designed to cover much more of the detailed content as the participants had a much more substantial background. In order to solidify the optical concepts, students were instructed to build group consensus on each phase of the activity instead of interacting directly with the facilitator as the authority figure. 

	The first step of the activity is a quick demonstration of a functioning system but with no detail given on what the parts are. The first task is to trace the light path and to identify all the parts by their technical names. Since the students have past optical experience as well as several recent AOSS lectures, the students are instructed to make a quantitatively accurate optical ray trace with estimates of focal lengths and beam diameters. In addition to creating a ray trace, students are asked to make a block diagram identifying major components to give the group a common language and agreement on the function of all AO workbench components. 
	
	After this parts identification step, the students are asked to take a quick break.  During this break, a few optics are removed. Depending on the skill of the students, typically two to four of the four beam expander lenses are removed and additional lenses of  different focal lengths are added to the activity. For the IfA Maui workbench, the DM was also on a tip-tilt-translation stage and can be mis-aligned to introduce registration as a concept. In addition to the removal of optics, extra post holes are mounted that are con-focal but non-conjugate. If the focal lengths of the lenses and post hole locations are chosen properly, some of these additional lenses and post holes will allow multiple confocal mountings and multiple beam diameters so that the students have multiple choices for creating beam expanders. 
		
	Once the disassembly and misalignment is complete and additional post holders have been mounted, the students are brought back and given the task to re-create the AO workbench using the lenses sitting beside it. The students have their original drawings and know what the original design was but the task is to build team consensus on why each lens goes in what post holder.  With multiple con-focal mounts, and multiple beam sizes students must realize that there is only one combination that both expands the beam to the proper size as well as re-images the aberrator onto the DM. The same is true for the beam expander re-imaging the DM onto the LLA. The task of building team consensus allows all members of the group to discuss and familiarize themselves with the technical terminology and for the reasoning behind the design choices.  Students must articulate why the beam diameters must be preserved.  The aberrator must be re-imaged onto the DM which requires understanding the concept of a relay. Even in collimated space, a specific wavefront shape can be relayed to another location and typically this concept is easily facilitated by suggesting teams draw diagrams using their measured focal lengths.

	In the final re-alignment phase of the activity, alignment techniques and good optics lab practices are facilitated. The nominal alignment procedure is to align the beam through the center of all non-powered optics. Lenses are mounted one-by-one making sure that each lens is mounted in the center and does not steer the beam. Once the re-alignment is completed, the software control system is demonstrated. The DM must be registered with the WFS. 

	This activity is nominally finished in 1-2 hours with AOSS students having significant background knowledge. The facilitation strategies depend somewhat on how much optics knowledge the group has, how easily consensus is built in the group and any social / cultural issues present in the group. As the activity is typically groups of four with only one workbench, equal sharing of materials and equitable respect for learner opinions and learning processes require some facilitation. Common facilitation issues are a dominant person doing all the talking and alignment, a disengaged learner, or assertive learners not questioning their assumptions.

\subsection{Systems Optimization}

	In both AMSC and AOSS activities there are opportunities to investigate or discuss different systems performance characteristics. The software and hardware control of optical components is critical in this type of investigation. The two different DM types strongly influence the types of activities. 
	
	The drum-head A-Optics DMs cannot take more complicated shapes easily but can compensate for a lot of defocus. The IrisAO DMs are segmented and can take complicated shapes rapidly. The IrisAO software also displays real-time statistics on which corrections are being applied to the DM in a given time interval. This gives instructors a tool to relate the system output to mathematical / theoretical concepts the students may have seen. 
	
	In the most basic implementation of the AO workbench, the aberrator is a simple low-powered lens used to defocus the system.  The DM can then re-focus the system. The quality of the correction can be investigated for lenses of different focal lengths to see that the performance degrades as more severe aberrations are introduced. This can be made quantitative for those groups with interest. 

	To investigate the system performance more thoroughly, one can introduce various aberrations and time-dependencies. These systems apply corrections to the DM from 10 to 60 times per second with a few different software parameters such as gain and aggressiveness. These system parameters must be chosen correctly to optimize performance. In the AMSC, students were given the opportunity to put various aberrations in the beam to see how good and how fast a correction the AO workbench could achieve. System performance characterization and optimization can be explored by varying these software parameters with different aberrators. In the courses for the new UHMC engineering technology BAS degree, a rotating aberrator has been introduced with a variable speed control so students can characterize the quality of the system correction as a function of aberration as well as system parameters. A detector has been placed at a re-imaged science camera focus to allow students to analyze digital data to quantify the quality of the image (Strehl ratio) and to measure system performance as a function of phase screen rotational velocity (in analogy to the wind speed). This is a larger activity in a remote sensing course to introduce students to data analysis, image calibration and processing, and systems optimization.

\section{Summary}

	One has many design choices when building an AO workbench but the various incarnations are useful for teaching many optical concepts, lab process skills and attitudes. The demonstrators have been used in undergraduate and graduate classrooms, summer-school laboratory activities for undergrads, graduates and professionals, as training devices and in public demonstrations and tours. There are many conceptual levels of activity from simple systems characterization up to optical conjugation and performance measurement. Students have the opportunity to hone their laboratory skills in optical alignment, systems investigation, documentation and operation. We have outlined a simple systems activity, multi-station investigations, engineering challenges and utilization of AO workbenches in multi-component classroom activities. These versatile systems are useful in the lab and classroom.

\acknowledgements 

	These systems were supported in many ways by many people. Instructors and staff of the AMSC included Mark Hoffman, Ryan Montgomery, Jess Johnson, Lisa Hunter, Lani Lebron, Hilary O'Bryan, John Pye, Jeff Kuhn, Elisabeth Reader, Mark Pitts, Mike Foley, Sarah Sonnett, and J.\ D.\ Armstrong. The AOSS lab activity part of a very large summer school and the AO workbench activity would have been impossible without Mark Ammons, Claire Max, Don Gavel, Scott Seagroves, Sylvana Yelda, Andrew Norton, Mike Fitzgerald and Katie Morzinski. We also with to thank Carl Kempf and Michael Helmbrecht at IrisAO for their support in constructing systems and mentoring students. The students themselves (James Ah Heong, Joe Curamen, Darcy Bibbs) assisted greatly. This material is based upon work supported by: the National Science Foundation (NSF) Science and Technology Center program through the Center for Adaptive Optics, managed by the University of California at Santa Cruz (UCSC) under cooperative agreement AST\#9876783; NSF AST\#0836053; NSF AST\#0850532; NSF AST\#0710699; Air Force Office of Scientific Research (via NSF AST\#0710699); UCSC Institute for Scientist \& Engineer Educators; University of Hawai`i.

\end{document}